# Superconductivity above 70 K observed in lutetium polyhydrides

Zhiwen Li[§,1,2], Xin He[§,1,2,3], Changling Zhang[§,1,2], Ke Lu[1,2], Baosen Min[1,2], Jun Zhang[1], Sijia Zhang[1], Jianfa Zhao[1], Luchuan Shi[1,2], Yi Peng[1,2], Shaomin Feng[1], Zheng Deng[1], Jing Song[1], Qingqing Liu[1], Xiancheng Wang*[1,2], Richeng Yu[1,2], Luhong Wang*[4,5], Yingzhe Li[5], Jay D. Bass[5], Vitali Prakapenka[6], Stella Chariton[6], Haozhe Liu[7], Changqing Jin*[1,2,3]

[1]*Beijing National Laboratory for Condensed Matter Physics, Institute of Physics, Chinese Academy of Sciences, Beijing 100190, China*
[2]*School of Physical Sciences, University of Chinese Academy of Sciences, Beijing 100190, China*
[3] *Songshan Lake Materials Laboratory, Dongguan 523808, China*
[4] *Shanghai Advanced Research in Physical Sciences, Shanghai 201203, China*
[5] *Department of Geology, University of Illinois at Urbana Champaign, Urbana, Illinois 61801, USA*
[6] *Center for Advanced Radiations Sources, University of Chicago, Chicago, Illinois 60637, USA*
[7] *Center for High Pressure Science & Technology Advanced Research, Beijing 100094, China*

The binary polyhydrides of heavy rare earth lutetium that shares a similar valence electron configuration to lanthanum have been experimentally discovered to be superconductive. The lutetium polyhydrides were successfully synthesized at high pressure and high temperature conditions using a diamond anvil cell in combinations with the *in-situ* high pressure laser heating technique. The resistance measurements as a function of temperature were performed at the same pressure of synthesis in order to study the transitions of superconductivity (SC). The superconducting transition with a maximum onset temperature ($T_c$) 71 K was observed at pressure of 218 GPa in the experiments. The $T_c$ decreased to 65 K when pressure was at 181 GPa. From the evolution of SC at applied magnetic fields, the upper critical field at zero temperature $\mu_0H_{c2}(0)$ was obtained to be ~36 Tesla. The *in-situ* high pressure X-ray diffraction experiments imply that the high $T_c$ SC should arise from the $Lu_4H_{23}$ phase with *Pm-3n* symmetry that forms a new type of hydrogen cage framework different from those reported for previous light rare earth polyhydride superconductors.

**Introduction**

As the lightest element, metallic hydrogen is expected to have a high Debye temperature and strong electron-phonon coupling which should lead to high temperature superconductivity (SC) based on Bardeen-Cooper-Schrieffer (BCS) theory[1]. However, the hydrogen metallization is hard to achieve since the predicted metallized pressure is beyond the capability of technologically accessible pressure to date[2]. To reduce the hydrogen metallization pressure, the polyhydride approach was proposed based on its chemical pre-compression effect[3, 4]. Sulfur hydrides of $SH_2$ or $SH_3$ were theoretically predicted to host high temperature SC with $T_c$ about 80 K at 160 GPa and 204 K at 200 GPa, respectively[5, 6]. Then high temperature SC was experimentally observed in the sulfur hydride system with critical temperature $T_c$ about 203 K under 155 GPa[7], which stimulated the investigation into SC in polyhydrides[8-11]. Following the discovery of a sulfur hydride superconductor, $LaH_{10}$ was synthesized and found to be superconducting with $T_c$ of 250 ~ 260 K at 170 ~ 200 GPa[12-15], $YH_9$ with $T_c$ of 243~262 K at 180-201 GPa and $YH_6$ with $T_c$ ~220 K at 183 GPa[16, 17], while $CaH_6$ with $T_c$ of ~210 K at 160-172 GPa[18, 19]. Besides those superconductors with $T_c$ exceeding 200 K, a handful of other polyhydride superconductors with moderate $T_c$ have been experimentally discovered as well, such as $ThH_{10}$ with the maximum $T_c$ of 161 K at 175 GPa[20] and 72 K at 200 GPa for $SnH_n$[21]. In addition, $ZrH_n$ ($T_c$ 71K)[22] were experimentally found to be the first IVB polyhydride of SC while $HfH_{14}$ ($T_c$ =83 K)[23] shows the highest $T_c$ SC of IVB polyhydride so far.

For the lanthanide polyhydrides, the SC seems to be related with the 4$f$ electrons since it was found that $T_c$ decrease with increasing 4$f$ electrons because of the spin scattering effects. $LaH_{10}$ has the highest $T_c$ while the maximum $T_c$ of $CeH_{10}$ goes down to 115 K[24]; it further decreases to 9 K and 5 K for $PrH_9$[25] and $NdH_9$[26], respectively. These experimental observations are in consistence with the theoretical calculations about the $f$ electrons dependence of $T_c$ for the light lanthanide polyhydrides[27]. However, for the heavy lanthanide of lutetium with a fully filled $f$ orbital, the $f$ electrons should contribute little to the electric density of state (DOS) near the Fermi level, and its effect on the SC of the polyhydride should be minimized. Lutetium and lanthanum have similar electronegativities and abilities to provide electrons to dissociate hydrogen molecules to atoms. Thus, lutetium polyhydride is expected to host high $T_c$ SC due to its fully filled $f$ shell. Here, we report the synthesis of $LuH_n$ and experimental discovery of SC in binary lutetium polyhydride. The SC was experimentally observed with $T_c$ = 71 K at 218 GPa. The structure investigation based on high pressure X diffractions with synchrotron suggests the superconducting transition is from the presence of the $Pm$-$3n$ $Lu_4H_{23}$ phase.

**Methods**

The lutetium polyhydrides were synthesized at high pressure and high temperature conditions using the diamond anvil cell (DAC) technique. The culet diameter of diamond anvils was about 50 μm which was beveled to 300 μm. T301 stainless was used as the gasket. The gasket was prepressed to ~10 μm thickness before being drilled with a hole of 300 μm in diameter, then filled with aluminum oxide. The aluminum oxide was densely pressed before further drilled to a hole of 40 μm in diameter serving as a sample chamber at high pressures. The ammonia borane (AB) was filled into the chamber to act as both the hydrogen source as well as the pressure transmitting medium. The inner Pt electrodes with a thickness of 0.5 μm were deposited on the surface of the anvil culet to serve as the inner electrodes, on which stacks a lutetium foil (99.9%) with 20 μm(L) * 20 μm(W) * 1 μm(T) size. The pressure was calibrated by the shift of Raman peak of the diamond anvil. The details are referred to in the ATHENA procedure reported in Ref.[28].

*In-situ* high pressure laser heating technique was adopted to generate high temperature. The wavelength of YAG laser is 1064 nm while the focused beam size is about 5 μm in diameter. The sample was laser heated at 2000 K for several minutes, with the temperature determined by fitting the black body irradiation spectra. The samples were quenched from high temperature while keeping the pressure unchanged. The high pressure electric conductivity experiments were performed in a MagLab system with temperatures varing from 300 K to 1.5 K and a magnetic field up to 5 Tesla. A Van der Pauw method was employed for the general high pressure resistance measurements[29, 30], with the applied electric current set to 1 mA.

The *in-situ* high pressure X-ray diffraction (XRD) experiments were performed at 13-IDD of Advanced Photon Source at the Argonne National Laboratory. The X-ray beam was focused down to ~3 μm in diameter with the wavelength of 0.3344 Å. The rhenium was used as the gasket to hold the high pressure samples, while a tiny Pt foil was loaded with samples into the pressure chamber. The samples were laser heated at 184 GPa to synthesize lutetium superhydrides. The pressure was kept unchanged during the diffraction experiments at room temperature. The pressure calibration was done by using both the equation of state from rhenium gasket material and the internal pressure marker Pt methods.

**Results & Discussions**

Fig. 1(a) shows the temperature dependence of resistance $R(T)$ for Sample A (Cell 1) synthesized and measured at 218 GPa, as well as for Sample B synthesized and measured at 181 GPa. The measurements were conducted in warming processes that gave a more homogeneous and better thermal equilibrium. The superconducting transition behaviors were observed with zero resistance achieved soon after the transition. The inset of Fig. 1(a) shows the derivative of resistance over temperature for Sample A, from which the onset superconducting $T_c$ of 71 K can be clearly determined with the upturn temperature. Fig. 1(b) displays the superconducting transition at different released pressures for Sample A. The inset of Fig. 1(b) shows the pressure dependence of $T_c$ for Sample A during releasing pressure. The $T_c$ monotonously decreased when pressure reduced to 193 GPa where the anvil became broken but with Tc trend comparable to that for Sample B synthesized at 181GPa.

To study the SC at magnetic field $H$, the measurements of electric transport at different $H$ were carried out as shown in Fig. 2(a), with the pressure of 213 GPa. The superconducting transition gradually shifted to low temperature when increasing $H$ that is in consistence with the SC nature. The dashed line marks the 90% of resistance relative to that of the normal state at the onset temperature. The upper critical magnetic fields $\mu_0 H_{c2}(0)$ at zero temperature were estimated using $T_c^{90\%}$ values that were determined by crosses between the dashed line and resistance curves at different $H$. Fig. 2(b) presents upper critical field $H_{c2}$ versus $T_c$. From the inset of Fig. 2(b), it can be seen that $H_{c2}(T)$ shows a linear behavior. The slope of $dH_c/dT$ was obtained to be -1.06 T/K upon linear fitting. Using the obtained $dH_c/dT$ slope value, the $\mu_0 H_{c2}(0)$ can be estimated to be ~48 T by the Werthamer-Helfand-Hohenberg (WHH) method with a formula of $\mu_0H_{c2}(T) = -0.69 \times dH_{c2}/dT|_{Tc} \times T_c$ by taking $T_c^{90\%} = 66$ K. In addition, $\mu_0H_{c2}(0)$ can be estimated by the Ginzburg Landau (GL) theory. Using the equation of $\mu_0H_{c2}(T) = \mu_0H_{c2}(0)(1-(T/T_c)^2)$, we carried out a fit as shown in Fig. 2(b) from which $\mu_0H_{c2}(0)$ was obtained to be ~36 T. The GL coherent length $\xi$ is calculated to be ~30 Å by the equation of $\mu_0H_{c2}(0) = \Phi_0/2\pi\xi^2$, where $\Phi_0 = 2.067 \times 10^{-15}$ Web is the magnetic flux quantum.

The *in-situ* high pressure X-ray diffraction experiments were carried out to investigate the possible superconducting phase. For the XRD experiments, Sample C has been synthesized under 185 GPa. Fig. 3(a) presents the typical XRD pattern. Besides the weak diffraction peaks arising from Re used as the gasket, the majority of the diffraction peaks can be indexed to two structures: one is a cubic *Pm-3n* lattice with $a = 5.3582$ Å and the other is a cubic *Fm-3m* lattice with $a = 3.7599$ Å. For the *Pm-3n* lattice, only $Re_4H_{23}$ lanthanide polyhydride was reported to host such a cubic structure with lattice constant $a = 5.86$ Å for $Eu_4H_{23}$ (at 130 GPa)[31] while with $a = 6.07$ Å for $La_4H_{23}$ (at 150 GPa)[32]. Those lattice parameters are comparable with what we observed for the *Pm-3n* lattice in our lutetium polyhydrides. Therefore the *Pm-3n* lattice here is proposed to be from the $Lu_4H_{23}$ phase in our samples.

Both lanthanide polyhydrides of $LnH_3$ and $LnH_{10}$ are well known to crystalize into a *Fm-3m* lattice at megabar pressures. However, $LnH_{10}$ usually crystalizes into a larger lattice than that for $LnH_3$. For example, the lattice parameter $a$ of $LaH_{10}$ is 5.22 Å at 140 GPa[32] while the lattice parameter $a$ for $LuH_3$ is 4.29 Å at 122 GPa[33]. Here the *Fm-3m* lattice constant of $a$ at 185 GPa is 12.3% smaller than that of $LuH_3$ at 122 GPa[33]. The lattice shrink suggests potentially another *Fm-3m* phase of lutetium hydride rather than $LuH_3$ or $LuH_{10}$ was synthesized in our experiments. In fact,

*Fm-3m* ScH was theoretically stable at 200 GPa[34], while YH can be experimentally obtained by heating YH$_3$ under 130 GPa and was reported to be a *Fm-3m* lattice with $a$ = 3.90 Å at 170 GPa[16], which is very close to the observed lattice constant of *Fm-3m* structure in our experiments. Thus it is suggested that the *Fm-3m* lattice is from LuH phase.

Therefore the crystalline structures of *Pm-3n* Eu$_4$H$_{23}$[31] and *Fm-3m* ScH[34] were adopted to be the initial models to perform the XRD structural refinements. The refinements smoothly converge with $R$wp = 11.6% and $R$p = 8.1%, indicating reasonableness of the structure models. The crystal structures of Lu$_4$H$_{23}$ and LuH are displayed in Fig. 3(b) and Fig. 3(c), respectively. In the *Pm-3n* Lu$_4$H$_{23}$ structure, there are two Wyckoff positions for Lu atoms: Lu1 (0, 0, 0) and Lu2 (0.25, 0, 0.5), which are surrounded by hydrogen atoms to form H$_{20}$ and H$_{24}$ cages, respectively. If the Wyckoff positions of the hydrogen atom in *Pm-3n* Eu$_4$H$_{23}$[31] are referred to for Lu$_4$H$_{23}$, the shortest H~H bond length in Lu$_4$H$_{23}$ is about 1.23 Å at 185 GPa that is within the range of H~H bond length from 1.0 to 1.5 Å for typical high $T_c$ superconducting polyhydrides[6, 9, 10, 34, 35]. For the *Fm-3m* LuH structure, Lu and H atoms are located at the fixed positions of Lu (0, 0, 0) and H (0.5, 0.5, 0.5) where the H atoms occupy the octahedral (O) interstitial sites of the Lu lattice and leave the tetrahedral (T) interstitial sites (0.25, 0.25, 0.25) empty. The shortest distance between adjacent H atoms is 2.65 Å, which is even longer than that in solid hydrogen at 15 GPa[36], implying that the electrons of H in the LuH structure tend to be localized.

In addition for the heavy lanthanide polyhydrides, *Immm* LuH$_8$ was theoretically predicted to be dynamically stable above 250 GPa and host SC with $T_c$ ~86 K at 300 GPa[27], and *Fm-3m* LuH$_3$ was experimentally observed to be SC with $T_c$ ~ 12 K at 122

GPa[33]. However, these previously reported superconducting phases could not be observed in XRD patterns in our sample. Therefore it is proposed that the observed SC should be from the presence of the *Pm*-3*n* Lu$_4$H$_{23}$ phase.

Up to now only three lanthanide polyhydrides of *Pm*-3*n* Re$_4$H$_{23}$ formula are experimentally reported, i.e., La$_4$H$_{23}$[32], Eu$_4$H$_{23}$[31] plus Lu$_4$H$_{23}$. No superconducting properties are studied for La$_4$H$_{23}$ although it is highly expected to be a high $T_c$ superconductor while Eu$_4$H$_{23}$ was predicted to be in the ferromagnetic ground state with a Curie temperature about 336 K[31]. Hence Lu$_4$H$_{23}$ is the first superconducting lanthanide polyhydride with *Pm*-3*n* structure. For the lanthanide polyhydrides with the same phase they should have comparable superconducting properties since the rare earth metals have the similar electronegativity and ability to provide electrons to dissociate hydrogen molecules to atoms. However, for the light lanthanide polyhydrides, the *f* electrons contribution to the DOS near the Fermi level would increase when increasing the number of *f* electrons, which weakens the electron phonon coupling and thus suppresses $T_c$[9, 25-27]. For the middle and heavy lanthanide polyhydrides, the local unpaired *f* electrons tend to generate magnetic order and thus are considered to be against SC, for example Eu$_4$H$_{23}$ with a magnetic ground state[31]. The lutetium polyhidrides are special in that the *f* shell of lutetium is fully filled so the DOS near Fermi level derived from *f* electrons should become small. Hence the effect of *f* electrons on SC is minimized in lutetium polyhydride.

Recently, nitrogen doped lutetium hydride of LuH$_{3-\delta}$N$_\varepsilon$ was claimed to show possible evidence of room temperature SC at near ambient pressure of 1 GPa, which was accompanied by peculiar color change from dark blue in the low pressure non SC phase across pink in the pressure range for the SC phase to bright red for another non

SC phase with further increasing pressure[37]. Based on the structure model in that paper, the shortest H~H distance for their LuH$_{3-\delta}$N$_\varepsilon$ is estimated to be ~2.17 Å. This is surprisingly large and almost twice that for YH$_9$ and LaH$_{10}$ whose $T_c$ are approaching room temperature[9, 34]. If the report was real, then what role does hydrogen play in the assumed near ambient SC in LuH$_{3-\delta}$N$_\varepsilon$? Shortly after the claim, a followup paper reported the very similar color change in LuH$_2$ without nitrogen doping[38], i.e., it transforms from dark blue to pink and then to bright red with increasing pressure in the same sequence as observed in the LuH$_{3-\delta}$N$_\varepsilon$ SC sample. However, non SC was observed within 7 GPa. These results implied that the claimed high temperature SC in LuH$_{3-\delta}$N$_\varepsilon$ seems irrelevant to the color change. Anyway, intensive doubts remain about the claim of ambient temperature SC in LuH$_{3-\delta}$N$_\varepsilon$

Conclusion

In summary, lutetium polyhydrides were successfully synthesized at high pressure and high temperature conditions. The SC was found with $T_c$ = 71 K at 218 GPa. The $\mu_0 H_{c2}(0)$ was estimated to be ~36 T from GL formula. The *in-situ* high pressure structural analysis suggests that the SC is likely from the *Pm*-3*n* Lu$_4$H$_{23}$ phase. This is another high $T_c$ superconductor of lanthanide polyhydride with a different type of hydrogen cage framework.

**Acknowledgments**: The works are supported by NSF, MOST & CAS of China through research projects (2018YFA0305700, 2021YFA1401800, XDB33010200). The *in-situ* high pressure X-ray experiments were performed at GeoSoilEnviroCARS (The University of Chicago, Sector 13), Advanced Photon Source (APS), Argonne National Laboratory. GeoSoilEnviroCARS is supported by the National Science Foundation – Earth Sciences (EAR – 1634415). This research used resources of the Advanced Photon Source, a U.S. Department of Energy (DOE) Office of Science User Facility operated for the DOE Office of Science by Argonne National Laboratory under Contract No. DE-AC02-06CH11357.

**Figure Captions**

**Figure 1**. The superconductivity measurements. (a) Temperature dependence of electric resistance for Sample A at 218 GPa and Sample B at 181 GPa of lutetium polyhydride (see Text for more details). The inset is the derivation of electric resistance over temperature to define the superconducting transition temperature wherein a $T_c$ onset is about 71 K for Sample A; (b) The superconducting transitions measured at different released pressures for Sample A. The inset shows the pressure dependence of $T_c$ during releasing pressure.

**Figure 2**. The superconducting parameters. (a) Temperature dependence of electric resistance measured at 213 GPa in different magnetic fields. (b) The upper critical magnetic field $\mu_0 H_{c2}(T)$ with $T_c^{90\%}$ being adopted. The red line is from the GL fitting. The inset shows the linear fitting results.

**Figure 3** Structure at high pressure. (a) The *in-situ* high pressure X-ray diffraction pattern measured at 185 GPa and the refinements. (b) and (c) are the crystal structures of *Pm*-3*n* $Eu_4H_{23}$ and *Fm*-3*m* LuH, respectively.

Fig. 1

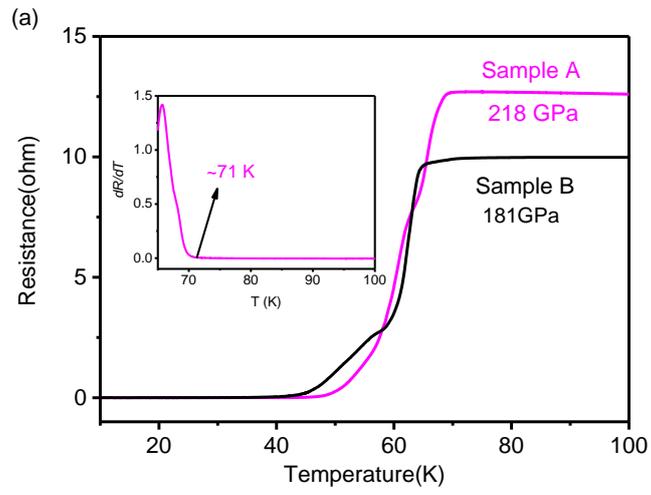

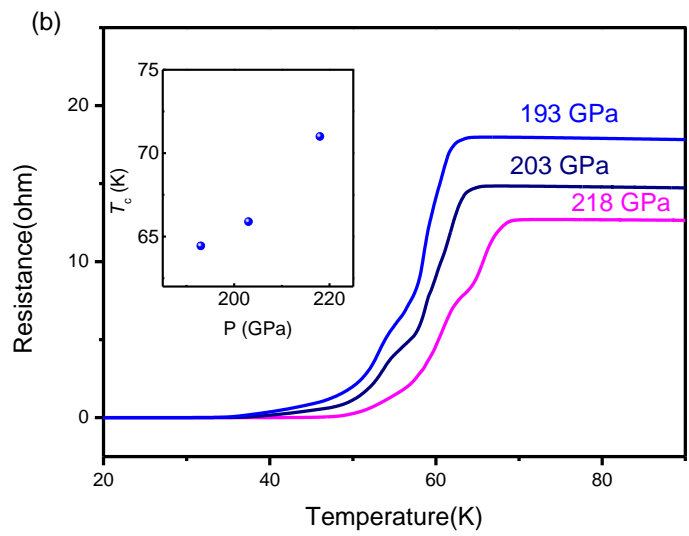

Fig. 2

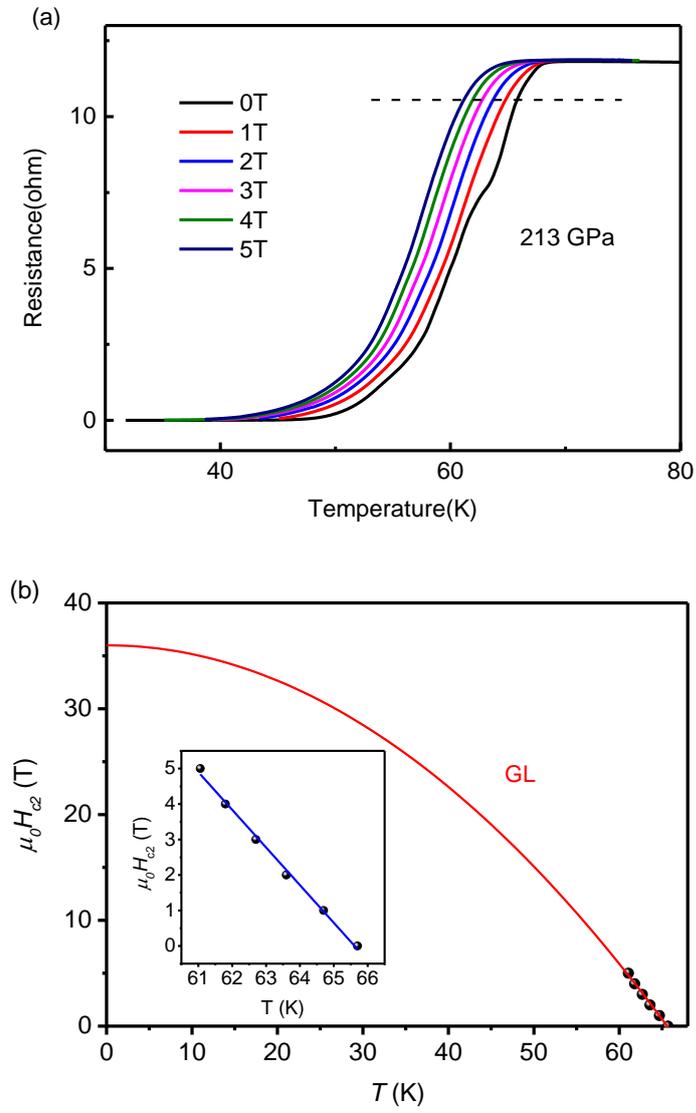

Fig. 3

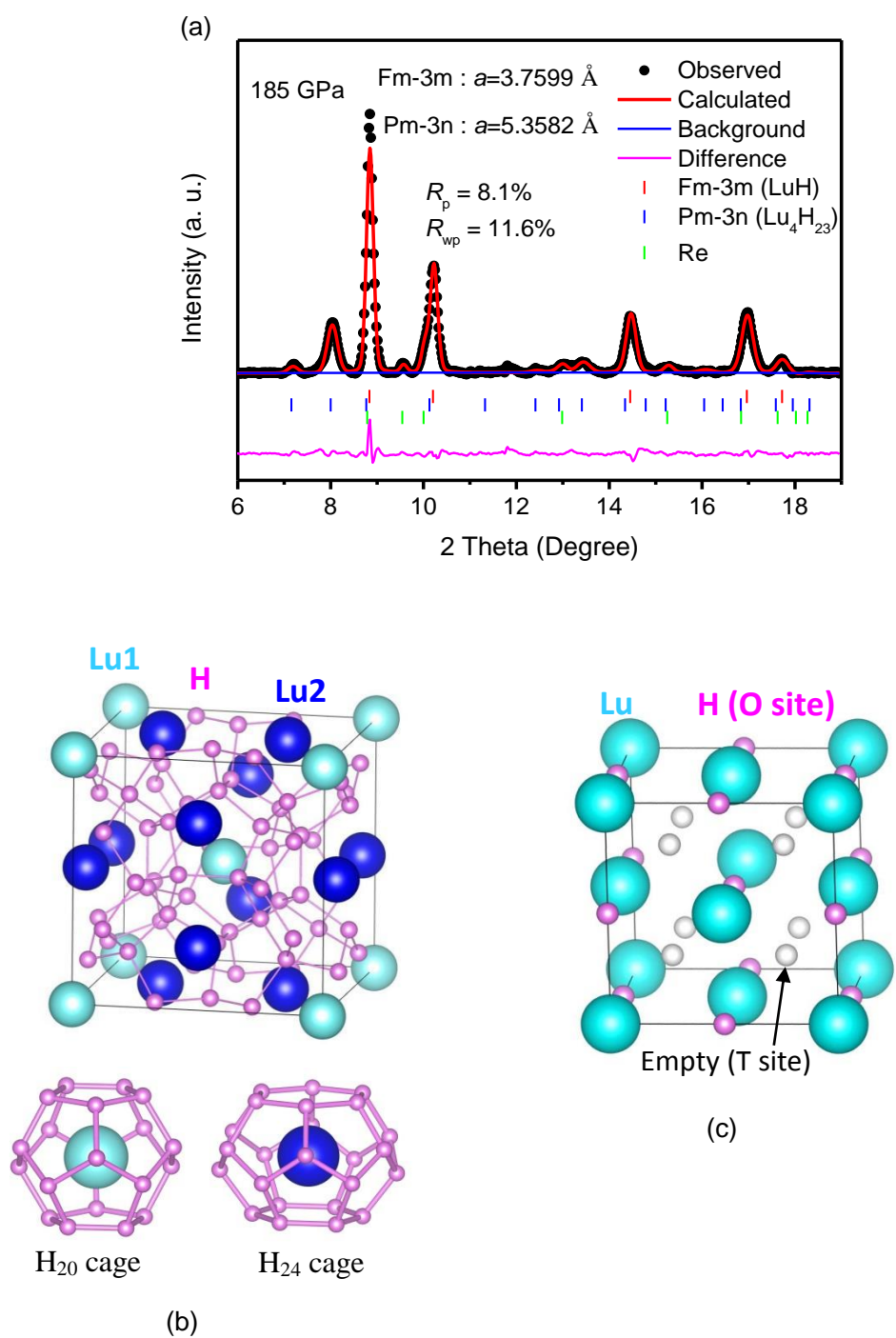

(a)

(b) H$_{20}$ cage   H$_{24}$ cage

(c)